\documentclass[]{pasj01}

\usepackage{graphicx}
\usepackage{epsfig}
\usepackage{natbib}    

\begin{document} 
\Received{}
\Accepted{}

\title{Giant Molecular Cloud Collisions as Triggers of Star Formation. VI. Collision-Induced Turbulence}

\author{Benjamin \textsc{Wu}\altaffilmark{1}}
\altaffiltext{1}{National Astronomical Observatory of Japan, Mitaka, Tokyo 181-8588, Japan}
\email{ben.wu@nao.ac.jp}
\author{Jonathan C. \textsc{Tan},\altaffilmark{2}\altaffilmark{3}}
\altaffiltext{2}{Department of Physics, University of Florida, Gainesville, FL 32611, USA}
\altaffiltext{3}{Department of Astronomy, University of Florida, Gainesville, FL 32611, USA}
\author{Fumitaka \textsc{Nakamura}\altaffilmark{1}}
\author{Duncan \textsc{Christie}\altaffilmark{3}}
\author{Qi \textsc{Li}\altaffilmark{3}}


\maketitle

\begin{abstract}
We investigate collisions between giant molecular clouds (GMCs) as
potential generators of their internal turbulence. Using
magnetohydrodynamic (MHD) simulations of self-gravitating, magnetized,
turbulent, GMCs, we compare kinematic and dynamic properties of dense
gas structures formed when such clouds collide compared to those that
form in non-colliding clouds as self-gravity overwhelms decaying
turbulence. We explore the nature of turbulence in these structures
via distribution functions of density, velocity dispersions, virial
parameters, and momentum injection. We find that the dense clumps
formed from GMC collisions have higher effective Mach number, greater
overall velocity dispersions, sustain near-virial equilibrium states
for longer times, and are the conduit for injection of turbulent
momentum into high density gas at high rates.
\end{abstract}

\section{Introduction}



Collisions between dense molecular clouds (MCs) within the
interstellar medium (ISM) have long been posited to explain various
star-forming regions observed in the Galaxy 
\citep[e.g.,][]{Loren_1976,Scoville_Sanders_Clemens_1986,Duarte-Cabral_ea_2011,Fukui_ea_2014}.
On a global galaxy scale, giant molecular cloud (GMC) collisions
driven by galactic orbital shear \citep{Gammie_ea_1991} may be a
dominant mode for the creation of star clusters and high-mass stars,
thus setting the star formation rates (SFRs) in circumnuclear
starbursts and disk galaxies \citep{Tan_2000}. These rates are characterized
by the dynamical Kennicutt-Schmidt relation
\citep{Kennicutt_1998,Leroy_ea_2008,Suwannajak_ea_2014}, $\Sigma_{\rm
  SFR} \propto \Sigma_{\rm gas}\Omega$, where $\Sigma_{\rm SFR}$ is
the SFR surface density, $\Sigma_{\rm gas}$ is the total gas mass
surface density, and $\Omega$ is the orbital angular frequency. For a
flat rotation curve disk, $\Sigma_{\rm SFR} \propto \Sigma_{\rm
  gas}/t_{\rm orb}$, where $t_{\rm orb}$ is the orbital time. The average 
GMC collision time in a marginally self-gravitating disk with a significant 
fraction of its gas mass in GMCs is expected to be a small, approximately 
constant fraction of the orbital time, i.e., $t_{\rm coll}\sim 0.2 t_{\rm orb}$ 
\citep{Tan_2000},
depending somewhat on the GMC mass function. This relation was
verified approximately by \citet{Tasker_Tan_2009} and \citet{Tan_Shaske_Van_Loo_2013}, 
with more detailed investigation by \citet{Li_ea_2017arXiv} 
\citep[see also, e.g.,][]{Dobbs_2008,Fujimoto_ea_2014,Dobbs_ea_2015}.
The mechanism of shear-driven GMC collisions as the rate limiting step for creating
star-forming clumps naturally connects the scales of star cluster
formation, i.e., $\sim1-10\:$pc, with those of global galactic disks,
i.e., $\sim1-10\:$kpc.

GMCs are observed to have highly supersonic velocity dispersions,
indicative of turbulence playing a large role in their structure and
evolution
\citep[e.g.,][]{Larson_1981,McKee_Ostriker_2007}.
This is supported by
parsec-scale numerical simulations of supersonic magnetized turbulence
\citep[e.g.,][]{Federrath_Klessen_2012,Padoan_ea_2014}.  However,
simulations also show that cloud-scale turbulent modes decay within
$\sim$1 dynamical time, $t_{\rm dyn}\sim R_{c}/\sigma_{c} \sim t_{\rm
  ff}$, where $R_{c}$ is the cloud radius and $\sigma_{c}$ is the
internal 1D velocity dispersion
\citep{Stone_ea_1998,MacLow_ea_1998,MacLow_1999}. Thus,
mechanisms are sought to explain the replenishment of turbulence in
GMCs. Feedback from star formation has been proposed, e.g., from
ionization fronts and/or supernovae 
\citep[e.g.,][]{Joung_MacLow_2006,Goldbaum_ea_2011,Koertgen_ea_2016}. 
However, it is unclear
if realistic models of feedback are efficient enough to power the
observed levels of turbulence in GMCs, especially given the
``impedance mismatch'' of coupling supernova feedback, which mostly
permeates low-density phases of the ISM, with the dense,
over-pressured conditions of GMCs. Furthermore, there is no clear
observational evidence that the degree of turbulence, e.g., as
measured via the virial parameter \citep[][see below]{Bertoldi_McKee_1992}.
is greater for GMCs with active star formation compared to those
without.

Frequent collisions between GMCs may be a source of stochastic
turbulent energy and ``turbulent momentum'' injection within the
clouds \citep{Tan_2000,Tan_Shaske_Van_Loo_2013,Li_ea_2017arXiv}. Much of the energy
will be radiated away in post-shock cooling layers, so we focus on
``turbulent momentum'', by which we are referring to the momentum
associated with internal turbulence in the GMC(s). One goal of this
present study is to relate turbulent momentum injection to that of the
initial momenta of the two GMCs, in their center-of-mass frame. Thus
we will investigate the effect that GMC collisions have on turbulence,
by following the momentum injected in interactions at the GMC
scale down to pre-stellar clumps and filaments.  

This work forms part of a series of papers investigating the nature of
collisions between magnetized GMCs. Paper I \citep{Wu_ea_2015}
explored an idealized 2-D parameter space for GMC collisions in ideal
MHD and introduced photo-dissociation region (PDR)-derived 
heating/cooling functions.  Paper II
\citep{Wu_ea_2017a} expanded the study to 3D and introduced supersonic
turbulence within the GMCs.  Paper III \citep{Wu_ea_2017b} developed
sub-grid star formation models and Paper IV \citep{Christie_ea_2017arXiv}
implemented ambipolar diffusion.  Paper V \citep{Bisbas_ea_2017arXiv}
investigated observational signatures via 
PDR and radiative transfer modeling.

Below, Section \ref{sec:model} describes our numerical model.  Results
are presented in Section \ref{sec:results}, including analysis of
morphology, density distributions, kinematics, dynamics, and turbulent
momentum.  We discuss our conclusions in Section
\ref{sec:conclusions}.

\section{Numerical Model}
\label{sec:model}

\subsection{MHD Simulations of GMC Collisions}

We perform our analysis on the fiducial ideal-MHD non-colliding and
colliding GMC simulations of Paper IV. These GMC models include
self-gravity, supersonic turbulence, PDR-based heating/cooling, and
are simulated with ideal MHD. Note that we are primarily focused on
the turbulent properties of the gas and thus do not include star
formation (introduced in Paper III) or ambipolar diffusion (introduced
in Paper IV) in this particular analysis.

Our model comprises a $(128\:{\rm pc})^{3}$ domain containing two
GMCs, initialized as uniform spheres of radius $R_{\rm GMC} =
20.0$~pc, Hydrogen number densities of $n_{\rm H,GMC} = 100\:{\rm
  cm^{-3}}$, and individual masses of $M_{\rm GMC} = 9.3 \times
10^4\:M_\odot$. The GMCs are initially nearly touching, with the
center of ``GMC 1'' at $(x,y,z)=(-R_{\rm GMC}, 0, 0)$ and ``GMC 2'' at
$(R_{\rm GMC}, 0, b)$, where the impact parameter $b=0.5\:R_{\rm
  GMC}$.  The clouds are embedded in an ambient medium of $n_{\rm H,0}
= 10\:{\rm cm^{-3}}$ representing an atomic cold neutral medium (CNM).
In the colliding model, both the CNM and GMCs are converging with a
relative velocity of $v_{\rm rel}=10\:{\rm km\:s^{-1}}$. In the
non-colliding model, $v_{\rm rel}=0\:{\rm km\:s^{-1}}$.

The entire domain is magnetized, initialized with a uniform magnetic
field of $|\textbf{B}|=10\:{\rm \mu G}$ at an angle
$\theta=60^{\circ}$ with respect to the collision axis.  The GMCs are
moderately magnetically supercritical, each with a dimensionless
mass-to-flux ratio $\lambda_{\rm GMC}=(M/\Phi)/(1/(2\pi
G^{1/2}))=4.3$. 
Equilibrium temperatures of GMC gas are $\sim 15$~K,
corresponding to a thermal-to-magnetic pressure ratio 
$\beta=8 \pi c_{s}^2 \rho_{0} / B^{2}=0.015$.

Within the GMCs, gas is initialized with a supersonic turbulent
velocity field that is purely solenoidal, following $v^{2}_{k} \propto
k^{-4}$, where $k=\pi / L$ is the wavenumber for an eddy diameter
normalized to the simulation box length, $L$.  The strength of the
initial turbulence is moderately super-virial, with a 1D velocity
dispersion of $\sigma_{v}=5.2\:{\rm km\:s^{-1}}$ corresponding to
sonic Mach number $\mathcal{M_{s}}\equiv \sigma/c_{s}=23$ (for $T=
15\:{\rm K}$).  The initial virial parameter is $\alpha_{\rm
  vir}\equiv 5 \sigma^{2}R / GM=6.8$.  This turbulence is not
artificially driven and would decay within a few dynamical times.

In this work, we follow the evolution to 4.5 Myr. The freefall time
based on the initial densities of the GMCs is $t_{\rm ff}=\sqrt{3\pi /
  32G\rho}\simeq 4.35$~Myr, but $t_{\rm ff}$ for denser, substructures
that form from compression due to the collision or local turbulence can be 
	much shorter.

The simulation has been performed using
\texttt{Enzo}\footnote{http://enzo-project.org (v2.5)}, an MHD
adaptive mesh refinement (AMR) code \citep{Bryan_ea_2014}.  The MHD
equations are solved using a MUSCL-like Godonov scheme \citep{vanLeer_1977}
while the $\nabla \cdot \bf{B}=0$ solenoidal constraint of the
magnetic field is maintained via Dedner-based hyperbolic divergence
cleaning \citep{Dedner_ea_2002,Wang_Abel_2008}. The variables are
reconstructed using a simple piecewise linear reconstruction method
(PLM), while the Riemann problem is solved via the Harten-Lax-van Leer
with Discontinuities (HLLD) method.

The simulation domain is realized with a top level root grid of
$128^{3}$ with 4 additional levels of AMR. 
The decision to refine to a finer level is based on the desire to
resolve the local Jeans length \citep{Truelove_ea_1997}. We require at
least eight cells of resolution, which is twice that recommended by
\citet{Truelove_ea_1997}, but lower than that used by, e.g., \citet{Federrath_ea_2011}. 
Our models have an effective maximum resolution of $2048^{3}$
corresponding to a minimum spatial cell size of $\Delta x =
0.0625\:{\rm pc}$. However, note that with the resolution implemented
here we will not fully satisfy the Truelove condition over the range
of densities we simulate, in particular at high densities where the
gas cools to $\sim10\:$K or less. We note that, since the GMCs are
partially supported by magnetic fields, the precise refinement
condition is likely to depend on the magneto-Jeans length, which will
be less stringent than that of resolving the Jeans length. Still, our
focus here is not to fully resolve fragmentation, but
rather to describe bulk properties of the turbulence in the clouds.

\subsection{Connected Extractions}

Much of our investigation involves analysis of dense sub-structures
within the GMCs.  To isolate these regions, we execute a clump-finding
algorithm based on density contours to identify topologically
disconnected structures within the entire domain.  The largest
contiguous sub-structure, by volume, in which all contained cells have
densities above a certain threshold is extracted. These density
thresholds are chosen to be $n_{\rm H}=10^{2}$, $10^{3}$, $10^{4}$,
and $10^{5}\:{\rm cm^{-3}}$, where each successive extraction is
contained within the lower density threshold region.  These connected
extractions (CEs) are denoted as ${\rm CE_{2}^{nc}}$, ${\rm
  CE_{3}^{nc}}$, ${\rm CE_{4}^{nc}}$, and ${\rm CE_{5}^{nc}}$ for the
non-colliding model and ${\rm CE_{2}^{c}}$, ${\rm CE_{3}^{c}}$, ${\rm
  CE_{4}^{c}}$, and ${\rm CE_{5}^{c}}$ for the colliding model,
respectively.  Occasionally, we omit the superscript or subscript when
describing CEs for more general cases.  Note that this method differs
from the connected extractions performed in Paper II, which were
defined via a $^{13}{\rm CO}(J=1-0)$ intensity threshold connected
within position-velocity space.

\section{Results}
\label{sec:results}

From the simulations, we explore parameters that describe specific
aspects of turbulence within the clouds and clumps.  For each study,
we directly compare non-colliding GMCs with those undergoing a
collision and analyze global properties as well as those of the CEs.
We investigate the morphology (Section \ref{sec:morphology}), volume
density PDFs (Section \ref{sec:pdf}), kinematics (i.e., velocity
dispersion) (Section \ref{sec:kinematics}), dynamics (i.e., virial
analysis) (Section \ref{sec:dynamics}), and turbulent momentum
(Section \ref{sec:pdot}).

\subsection{Morphology}
\label{sec:morphology}

\begin{figure*}[htb]
	\begin{center}
		\includegraphics[width=2\columnwidth]{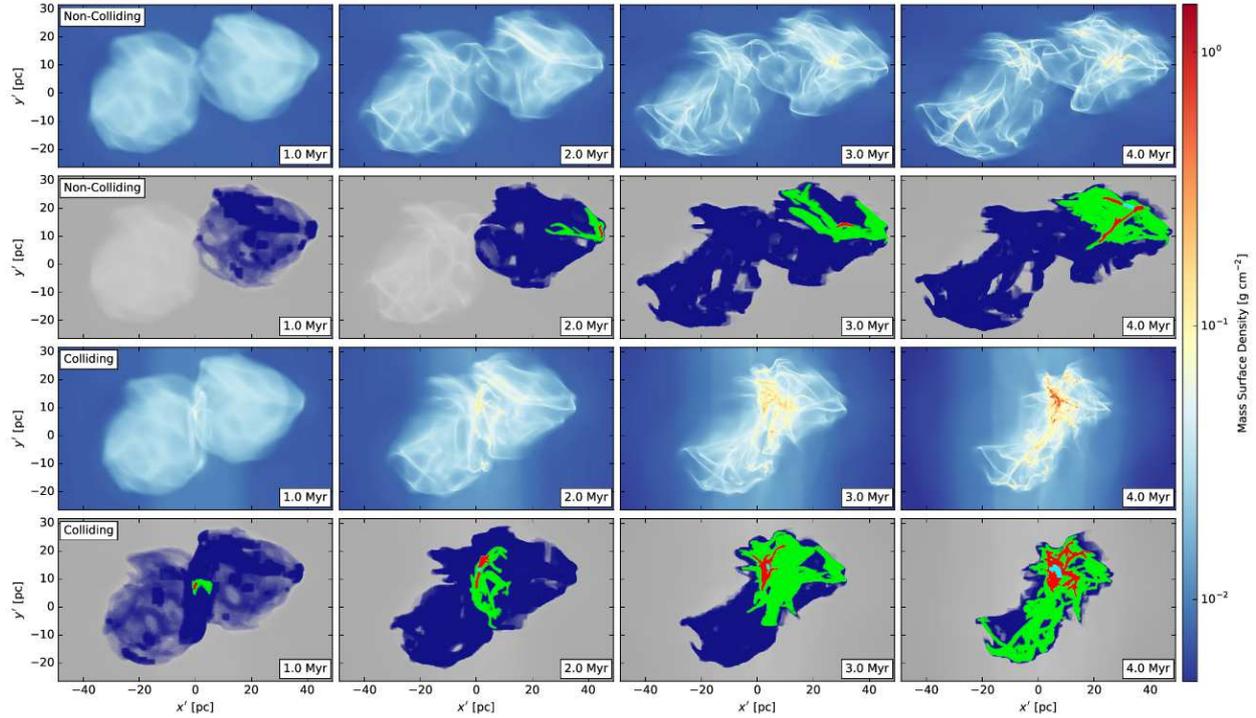} 
	\end{center}
	\caption{
		Row 1: Time evolution of projected mass surface density for the
		non-colliding model. Row 2: Connected Extractions (CEs), the largest
		spatially-connected sub-structures for each density threshold are
		shown plotted over the corresponding grayscale mass surface density
		map. Blue, green, red, and cyan correspond to ${\rm CE_{2}^{nc}}$,
		${\rm CE_{3}^{nc}}$, ${\rm CE_{4}^{nc}}$, and ${\rm CE_{5}^{nc}}$,
		respectively (see text). Rows 3-4: Same as rows 1-2 except for the
		colliding model. Snapshots at $t=1$, 2, 3, and 4 Myr are shown. The
		visualizations adopt the rotated coordinate system $(x',y',z')$ (i.e.,
		the simulation coordinates $(x,y,z)$ transformed by a $(\theta,
		\phi)=(\timeform{15D}, \timeform{15D})$ rotation used throughout this
		paper series.}\label{fig:proj}
\end{figure*}

Figure~\ref{fig:proj} shows the time evolution of the non-colliding
and colliding models. For each case, maps of mass surface density,
$\Sigma$, and the CEs are shown.

Evolution of the non-colliding case is dominated by the initial
turbulent velocity field and self-gravity.  A network of dispersed
filamentary structures develops over time, with slowly increasing
differentiation in mass surface density.  By $t=4\:{\rm Myr}$,
$\Sigma\approx 0.5\:{\rm g \: cm^{-2}}$ is seen in isolated regions.
Overall evolution is relatively quiescent and the spatial extent of
the original GMCs is roughly preserved.

${\rm CE_{2}^{nc}}$ contains most of the material from GMC 2 by
$t=1\:{\rm Myr}$ and, by 3 Myr, encompasses both GMCs.  ${\rm
  CE_{3}^{nc}}$ grows from a relatively isolated region near the
right-most edge of GMC 2 into a structure tens of parsecs in scale by
$4\:{\rm Myr}$, spanning much of GMC 2.  ${\rm CE_{4}^{nc}}$ retains a
highly filamentary morphology and generally occupies a small portion
GMC 2.  By $4\:{\rm Myr}$, it spans tens of parsecs and includes
multiple filaments, with ${\rm CE_{5}^{nc}}$ having formed near a
filament junction.

In contrast, the colliding case is dominated by two large-scale bulk
flows which compress both the GMCs and CNM.  This forms a centralized
conglomeration of filamentary structures with a primarily flattened
morphology oriented orthogonal to the collision axis.  Secondary
filaments extend outward from this region as well.  Large shocks sweep
up the gas in the growing collision region, eventually encompassing
all material from the originally separated GMCs.  Relative to the
non-colliding case, the collision forms dense structures with $\Sigma
> 0.5\:{\rm g \: cm^{-2}}$ at much earlier times and these are
predominantly clustered within the central main filament.

By $t=1\:{\rm Myr}$, the collision has bridged ${\rm CE_{2}^{c}}$ into
both GMCs, and it effectively traces the total combined GMC gas
throughout the simulation.  Additionally, ${\rm CE_{3}^{c}}$ and ${\rm
  CE_{4}^{c}}$ have formed at the GMC collision interface by this
time.  Over the next few Myrs, ${\rm CE_{3}^{c}}$ grows to encompass
most of GMC 2 and eventually the entire complex of high-density gas of
the primary filament.  ${\rm CE_{4}^{c}}$ and ${\rm CE_{5}^{c}}$
remain localized within the collisional interface and grow over time
as additional gas is fed into the region.

\subsection{Volume Density PDFs}
\label{sec:pdf}

\begin{figure*}[tb]
	\begin{center}
		\includegraphics[width=1\columnwidth]{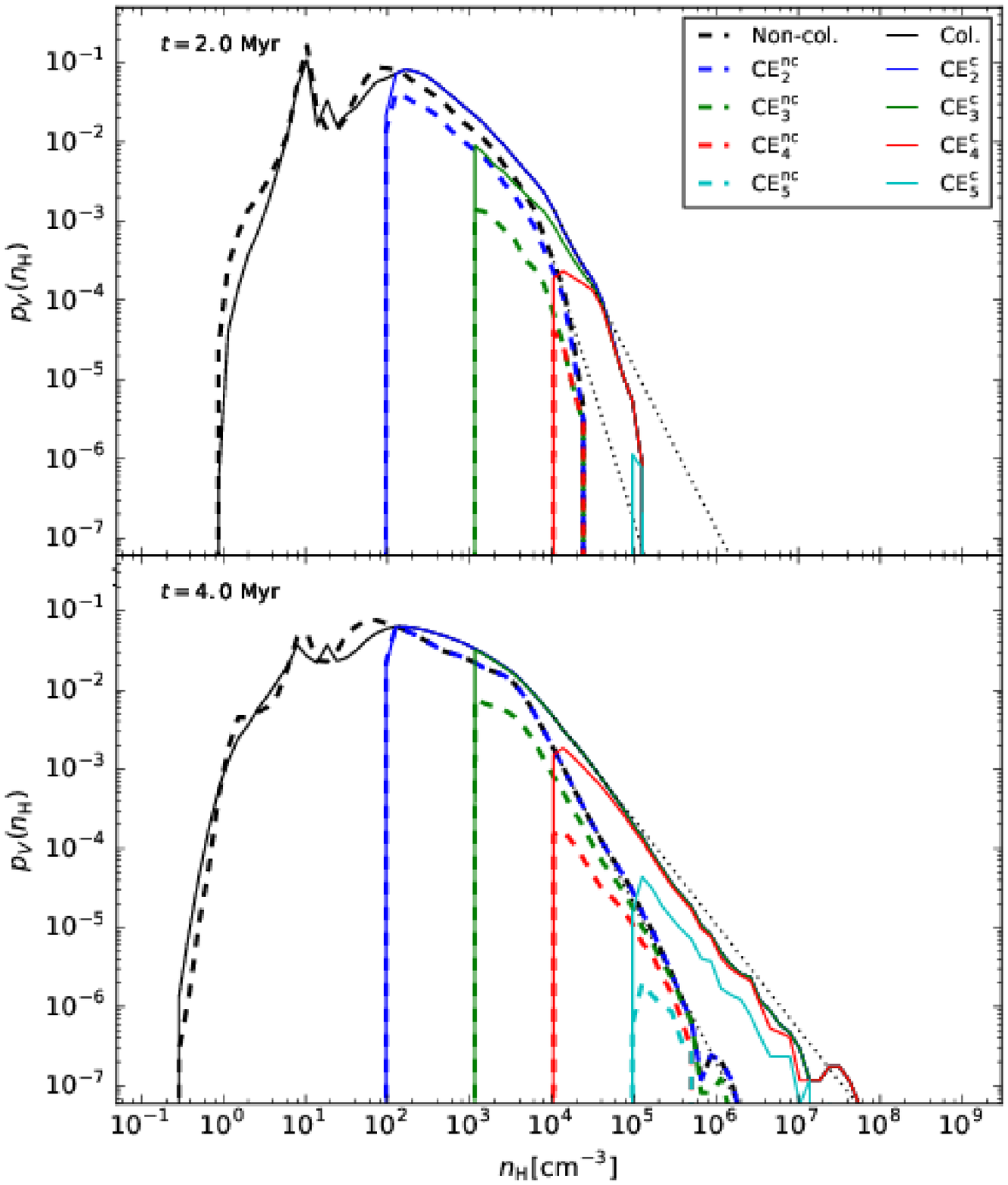}
		\includegraphics[width=1\columnwidth]{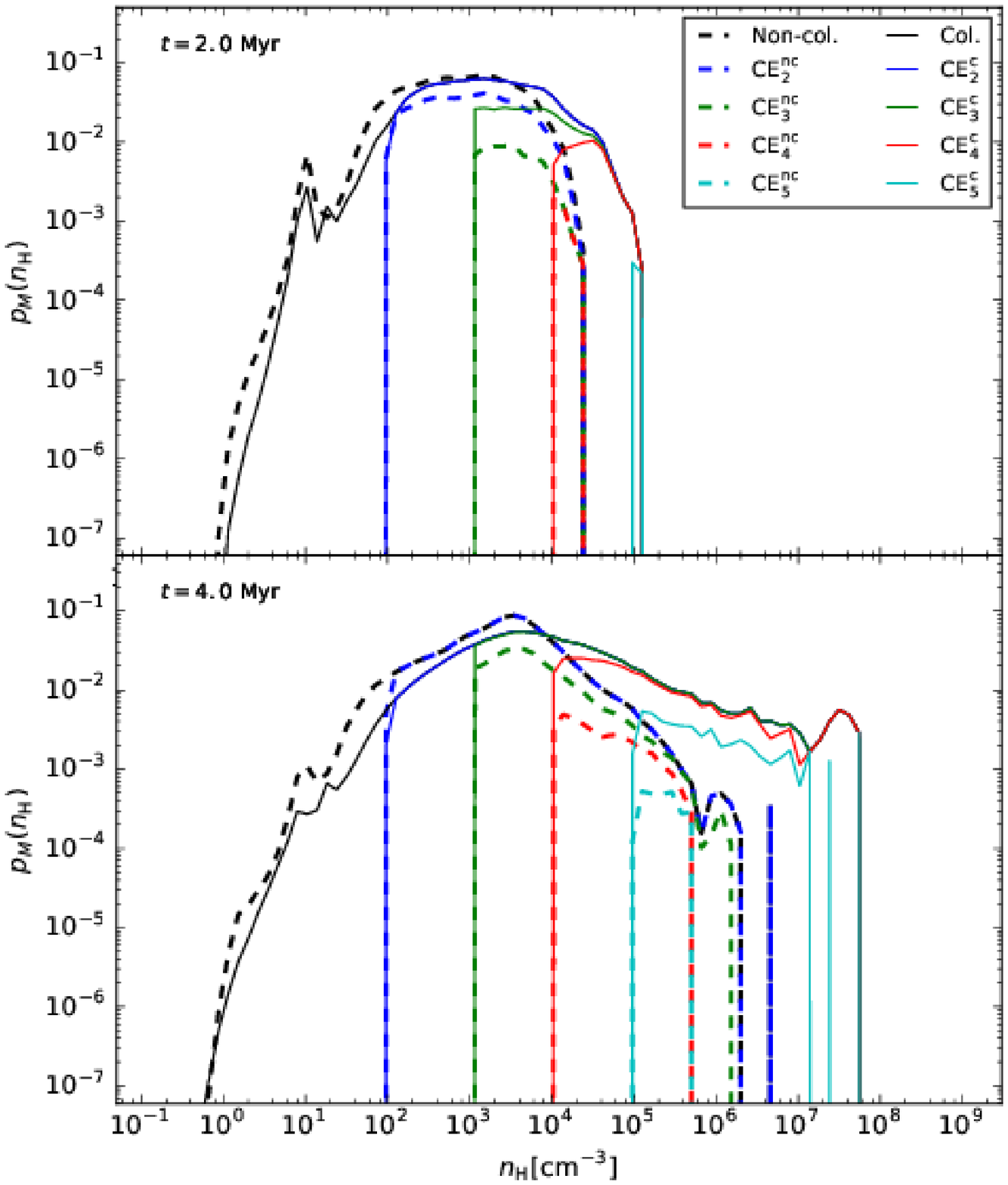}
	\end{center}
	\caption{
		(Left:) Volume-weighted PDFs at $t=2$ and 4 Myr
		of GMC evolution are shown. PDFs for the total gas (black), ${\rm
			CE_{2}}$ (blue), ${\rm CE_{3}}$ (green), ${\rm CE_{4}}$ (red), and
		${\rm CE_{5}}$ (cyan) are plotted. Dashed lines represent the
		non-colliding case, while solid lines represent the colliding
		case. Power laws are fit to each total gas PDF for $n_{\rm
			H}>10^{4}\:{\rm cm^{-3}}$. (Right:) Same as left, except for
		mass-weighted PDFs.  }\label{fig:pdf}
\end{figure*}

Probability distribution functions (PDFs) of gas have been used to
quantify various properties of supersonic turbulence. They can be
shaped by self-gravity, shocks, and magnetic fields 
\citep[see, e.g.,][]{Vazquez-Semadeni_1994,Padoan_ea_1997a,Padoan_ea_1997b,Kritsuk_ea_2007,Federrath_ea_2008,Price_2012,Collins_ea_2012,Burkhart_ea_2015}.
A notable
feature both in observations (of mass surface density, $\Sigma$) and
simulations (of volume density, $\rho$) is a lognormal distribution of
densities.  Paper II compared $\Sigma$-PDFs with an observed 
Infrared Dark Cloud IRDC from
\citet{Butler_ea_2014} and \citet{Lim_ea_2016arXiv} and found
consistency between the observed $\Sigma$-PDF and those formed in
colliding GMC simulations, though later stages ($t>4\:{\rm Myr}$) of
non-colliding GMCs were not ruled out.

A lognormal PDF takes the form:
\begin{equation}
{p_{V}(s) = \frac{1}{\sqrt{2\pi\sigma_{s}^2}} \exp \left( \frac{(s - \overline{s})^2} {2 \sigma_{s}^2} \right) d \ln \rho },
\end{equation}
where $s\equiv \ln(\rho/\rho_{0})$, and the mean ($\overline{s}$) and
variance ($\sigma_{s}$) of $\ln \rho$ are related by
$\overline{s}=-\sigma_{s}^2/2$ \citep{Vazquez-Semadeni_1994}.

The turbulent sonic Mach number, $\mathcal{M}$, has been shown in
simulations to be related to the standard deviation of the logarithm
of density, $\sigma_{s}$, via
\begin{equation}
{\sigma_{s} = \sqrt{\ln(1 + b^{2} \mathcal{M}^2)}}.
\end{equation}
The constant $b$, often referred to as the turbulence driving
parameter, depends on the turbulent driving modes of the acceleration
field.  From numerical simulations of periodic boxes of driven
turbulence, $b=1/3$ for fully solenoidal (divergence-free) forcing
\citep{Padoan_ea_1997b,Kritsuk_ea_2007,Federrath_ea_2008,Federrath_ea_2010},
while $b=1$ for fully compressive (curl-free)
forcing \citep{Federrath_ea_2008,Federrath_ea_2010}.

Figure~\ref{fig:pdf} shows volume-weighted and mass-weighted density
PDFs at $t=2$ and 4 Myr, normalized to total gas fraction.  PDFs for
non-colliding and colliding clouds, along with their respective CEs
are plotted together for comparison.

The volume-weighted PDFs at 2 Myr show a more broadened distribution
for the colliding case.  Densities have also reached an order of
magnitude higher at this time, to $\sim10^{5}\:{\rm cm^{-3}}$.  The
non-colliding PDF is fit with $\sigma_{s}=1.06$ and the colliding PDF,
with $\sigma_{s}=1.14$.  
Mass-weighted sonic Mach numbers, directly calculated from the
	individual cell values, yield $\mathcal{M}=10.90$ for the non-colliding case and
	$\mathcal{M}=13.05$ for the colliding. 
Values of $b=0.13$ were found (i.e., relatively low values, but not indicative 
	of any particular turbulent driving mode.)
The collision also produces CEs that occupy a higher
fraction of total gas at their given densities.  This indicates that
most of the high-density gas in a collision is contained within
connected structures concentrated at the collisional interface.  In
the non-colliding case, high-density gas is decentralized. This trend
is seen throughout all of the PDFs.

As the simulations progress to 4 Myr, the PDFs broaden in both models.
They show similar behavior at low-densities, but the collision forms a
higher-end tail, reaching almost two orders of magnitude greater
densities, to $\sim 10^{8}\:{\rm cm^{-3}}$.  These PDFs are fit with
$\sigma_{s}=1.42$ for the non-colliding case and $\sigma_{s}=1.63$ for
the colliding case.  Notably, the Mach number for the non-colliding
case has decreased to $\mathcal{M}=9.73$ while for the colliding case
it has instead increased to $\mathcal{M}=15.22$.  This may be indicative
of the collision inducing increased turbulence into the GMC-GMC system.  
In this time, the turbulent driving mode $b$ has
increased to 0.26 and 0.24 for the non-colliding and colliding models,
respectively. 
We note, with caveats that our boundary conditions are not those 
of a periodic box and that the precise effects of the non-turbulent ambient 
medium are unaccounted for, our measured values of $b$ are close to those 
seen in purely solenoidal driving simulations. Additionally, these have 
increased relative to earlier times, perhaps trending toward more 
compressive modes as gravitational collapse begins to dominate.
The similarity of $b$ measured between isolated and colliding clouds 
indicates that this parameter is not strongly affected by cloud collisions in our
particular simulation set-up.

\begin{figure*}[htb]
	\begin{center}		
		\includegraphics[width=2\columnwidth]{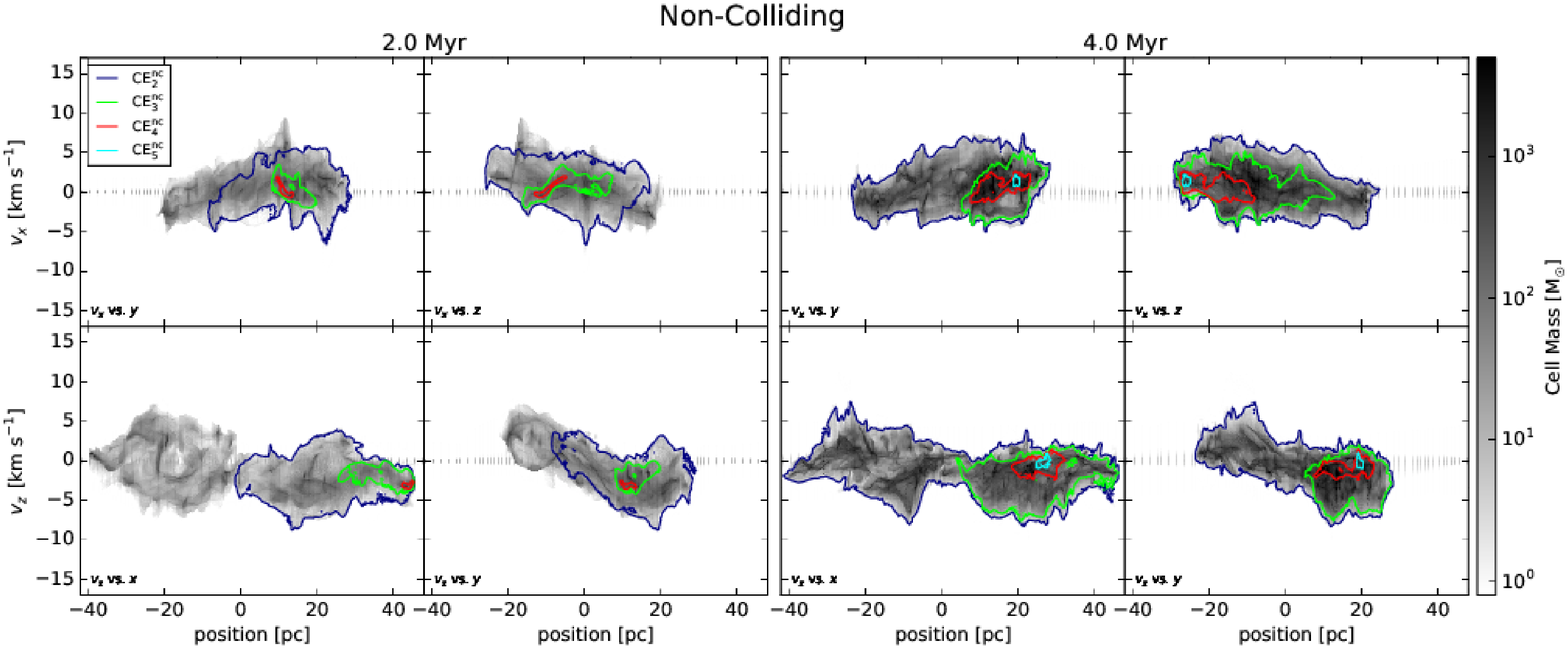}
		\includegraphics[width=2\columnwidth]{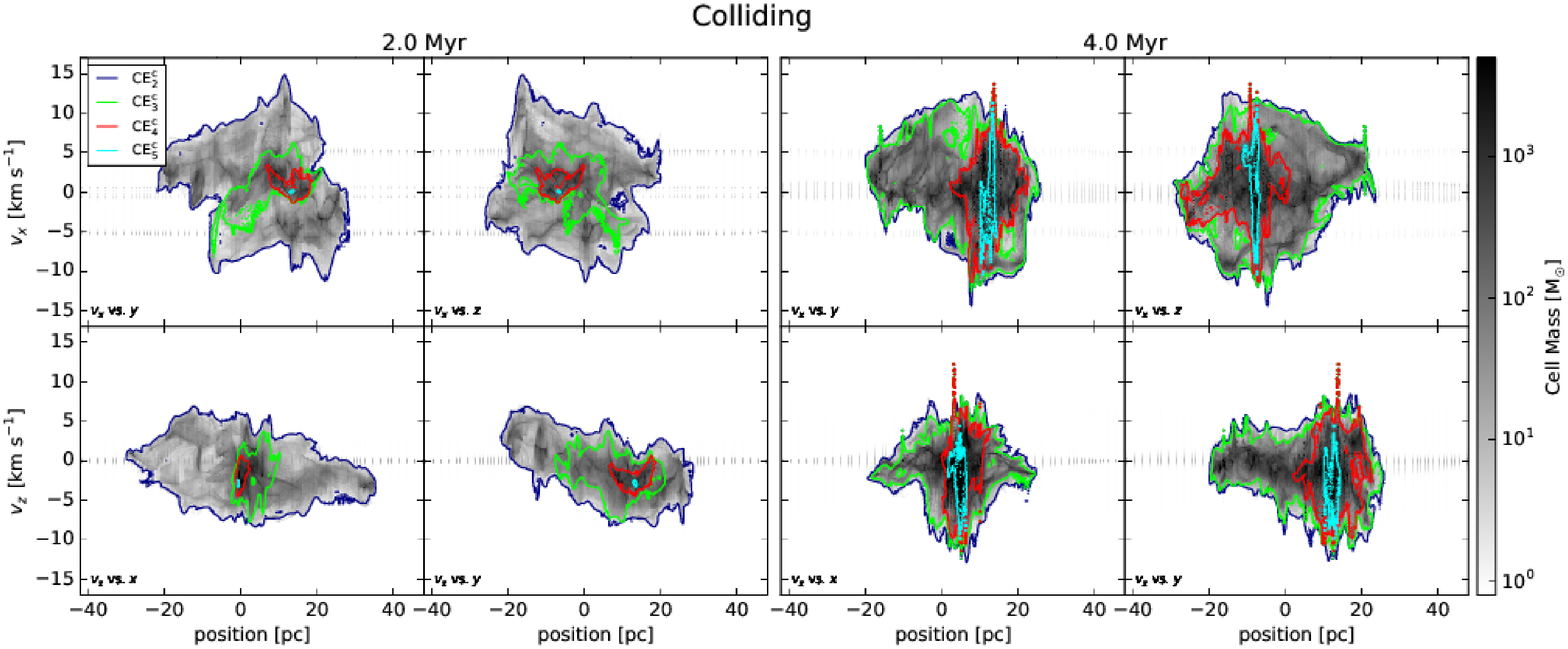}
	\end{center}
	\caption{
		Position-velocity diagrams for the non-colliding (top) and colliding
		(bottom) models. In each model, behavior at $t=2$ (left panels) and 4
		Myr (right panels) are shown. The sub-panels show line-of-sight
		velocities $v_{x}$ and $v_{z}$ versus orthogonal positions. The
		kinematic structure for the total gas is depicted as the grayscale
		colormap, while the CEs are denoted with correspondingly colored
		boundaries. 
		The edges of the shock in the ambient medium and the 
		associated central density enhancement can be seen in grey.
	}\label{fig:ppv}
\end{figure*}

Deviations from purely lognormal PDFs at high densities in the form of
power-law tails, i.e., $p(\rho)=a\rho^{-k}$, are often attributed to
gravitational collapse \citep{Klessen_2000,Vazquez-Semadeni_ea_2008}.
From isothermal self-gravitating supersonic turbulence simulations,
\citet{Kritsuk_ea_2011} found indices of −1.67 and −1.5 for intermediate
and high density thresholds, respectively, with flattening potentially
due to the onset of rotational support.  Self-gravitating MHD
simulations in \citet{Federrath_Klessen_2013} produced high-density tails
consistent with radial density profiles of power law index $k=-1.5$ to
-2.5.

We investigate power law fits to the high-density regimes ($n_{\rm
  H}>10^{4}\:{\rm cm^{-3}}$) of our PDFs.  At 2 Myr, the non-colliding
case has an index of $k=-3.35$, while the colliding case has $k=-2.03$.
At 4 Myr, the indices are $k=-2.00$ and $k=-1.32$ for the
non-colliding and colliding cases, respectively.  In both models, the
index becomes shallower as the clouds evolve, with the GMC collision
expediting this process.

The mass-weighted PDFs are presented as well in Fig.~\ref{fig:pdf}, 
and emphasize
differences between the non-colliding and colliding cases that grow
over time.  In subsequent sections, analysis of mass-weighted
quantities are commonly used.  These panels thus serve to more
intuitively display the relative contributions of gas at different
densities, particularly of the CEs.

\subsection{Kinematics}
\label{sec:kinematics}

\subsubsection{Position-velocity diagrams}
\label{sec:posvel}

\begin{figure*}[tb]
	\begin{center}
		\includegraphics[width=2\columnwidth]{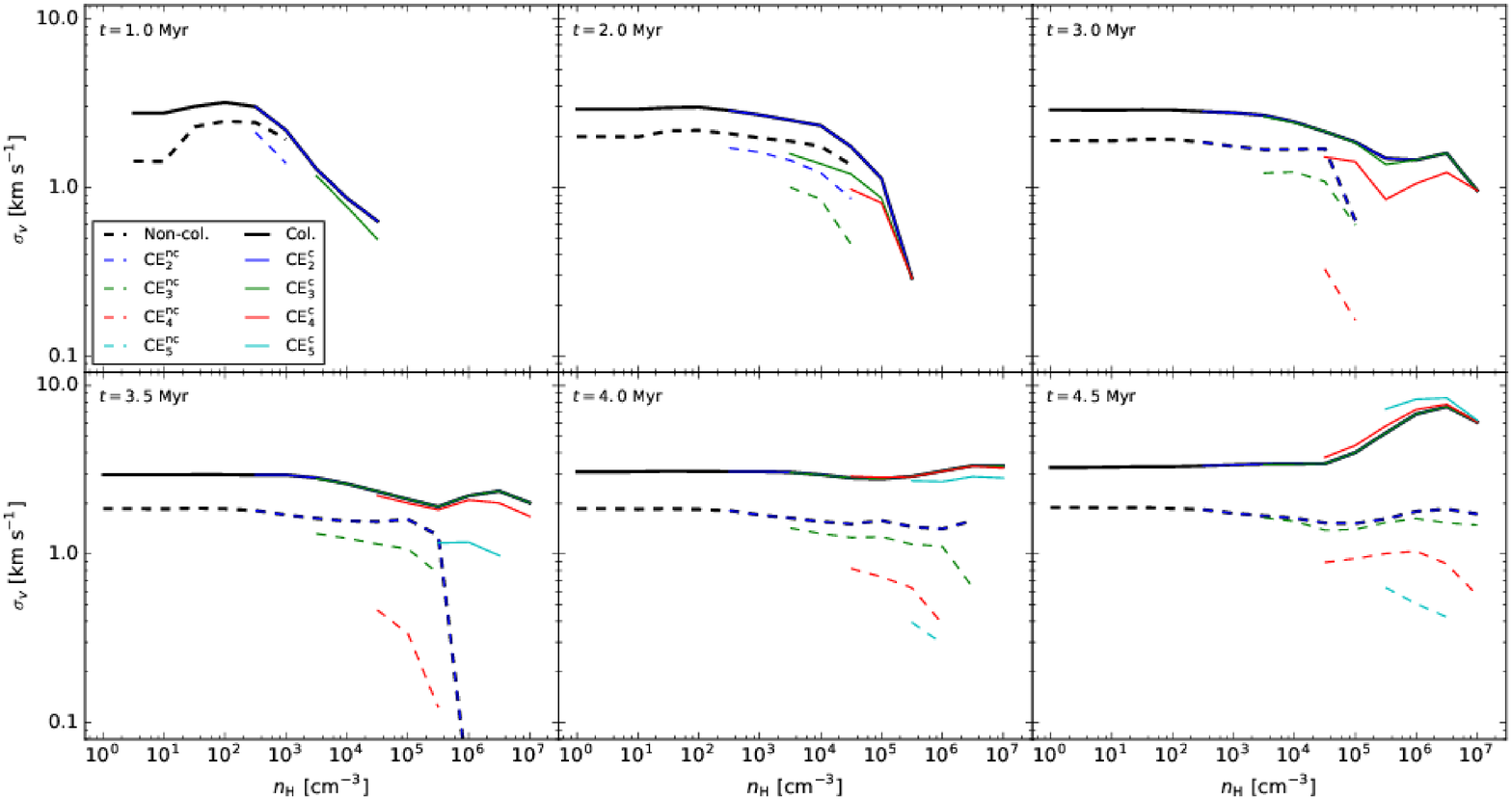} 
	\end{center}
	\caption{
		Velocity dispersions as functions of density. Each panel represents a
		different snapshot in time, from $t=1$ to 4.5 Myr. The non-colliding
		case is represented by dashed lines, while the colliding case is
		represented by solid lines. For each model, $\sigma_{v}$ (averaged
		over the $x$, $y$, and $z$ lines-of-sight) for the total gas content
		(black), $\rm{CE_{2}}$ (blue), $\rm{CE_{3}}$ (green), $\rm{CE_{4}}$
		(red), and $\rm{CE_{5}}$ (cyan) is plotted. }\label{fig:veldisp_bins}
\end{figure*}

We investigate the velocity structure of the entire clouds and
individual CEs via position-velocity ($p-v$)
diagrams. Observationally, this method is used to determine velocity
dispersion and velocity gradients of GMCs and IRDCs 
\citep[e.g.,][]{Hernandez_Tan_2015}.
Paper II investigated $p-v$ diagrams using
synthetic $^{13}{\rm CO}$-connected structures and found velocity
dispersions in colliding GMC simulations were higher than those of
non-colliding clouds by a factor of a few and were in agreement with
observed velocity dispersions of $\sim$ few ${\rm km\:s^{-1}}$
\citep{Hernandez_Tan_2015}.

Figure~\ref{fig:ppv} shows $p-v$ diagrams of the gas density in the
non-colliding and colliding simulations, measured along lines of sight
for the collision axis ($v_{x}$) and for a side view ($v_{z}$).
Results for $v_{y}$ are not displayed, but are in general similar to
those along $v_{z}$.

In the non-colliding model at 2 Myr, the overall gas distribution
shows turbulent kinematic structures that are generally within $\pm
5\:{\rm km\:s^{-1}}$.  The individual GMCs can be distinguished in the
$x$ direction.  ${\rm CE_{2}^{nc}}$, which traces GMC 2 at this time
(see Section~\ref{sec:morphology}), shows similar kinematic structure.
${\rm CE_{3}^{nc}}$ and ${\rm CE_{4}^{nc}}$ can also be seen, with
relatively narrow velocity spreads of at most a few ${\rm
  km\:s^{-1}}$.

At 2 Myr, the colliding model has formed a more spatially compact
structure and exhibits much stronger kinematic signatures.  Portions
of the GMCs exceed $\pm 10\:{\rm km\:s^{-1}}$, due to the bulk flow
velocity combined with initially turbulent gas. This gas has not yet
interacted with the incoming GMC.  ${\rm CE_{2}^{c}}$, which contains
the combined GMC gas, shows more disrupted structures with high
velocity dispersions.  The cloud collision can be clearly seen in the
$v_{x}$ vs. $y$ and $v_{x}$ vs. $z$ panels, with two large, opposing
velocity components bridged by an overdense intermediate-velocity
region.  It is within this collisional interface that the primary
higher-density structures--${\rm CE_{3}^{c}}$, ${\rm CE_{4}^{c}}$, and
${\rm CE_{5}^{c}}$--form.  Compared with the non-colliding model, the
velocity dispersion is much higher, although ${\rm CE^{c}}$s at
successively higher densities similarly form with lower relative
velocity dispersions.  The cloud collision is less apparent along
$v_{z}$, but higher velocity dispersions relative to the non-colliding 
case can be seen at the interface.

At 4 Myr, $p-v$ diagrams for the non-colliding model reveal the
presence of higher density structures via the total projected cell
mass values, but overall velocity dispersion is similar.  ${\rm
  CE_{2}^{nc}}$ now includes both GMCs, and ${\rm CE_{3}^{nc}}$ tracks
the majority of GMC 2.  ${\rm CE_{4}^{nc}}$ and ${\rm CE_{5}^{nc}}$,
which have now formed, contain gas with relatively low velocity
dispersions.

The colliding case at 4 Myr reveals very different behavior,
exhibiting much higher velocities and more compact gas structures.
CEs at this stage of evolution now include gas with much higher
velocity dispersions, 
with higher-density CEs showing large amounts of disruption.  Compared
to the colliding case at 2 Myr, the gas kinematics are now dominated
by post-collisional gas located in the central structure rather than
the pre-collisional gas from the original GMCs.

\subsubsection{Velocity dispersion}
\label{sec:veldisp}

\begin{figure*}[tb]
	\begin{center}
		\includegraphics[width=2\columnwidth]{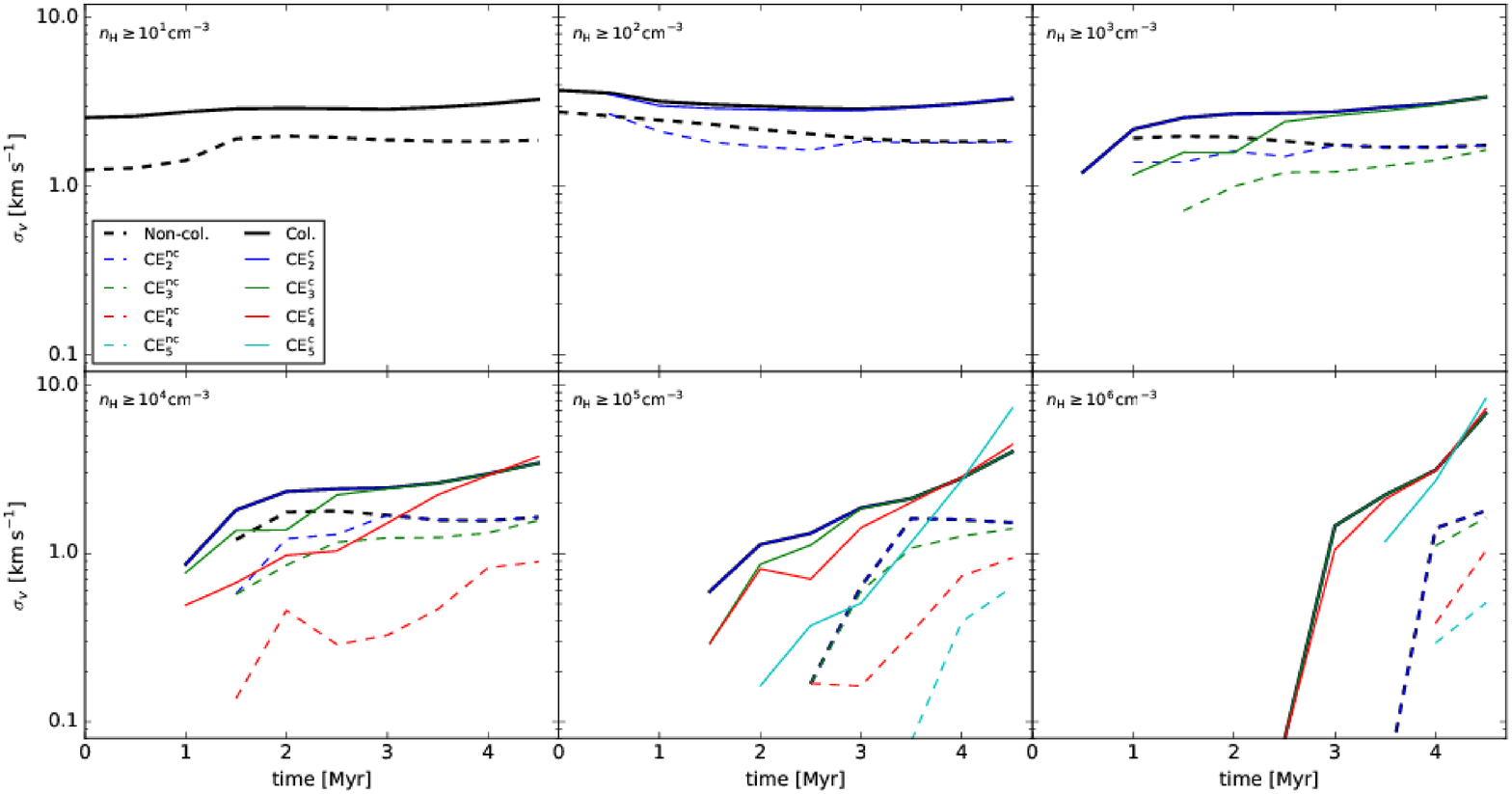}
	\end{center}
	\caption{
		Velocity dispersion as a function of time. Each panel displays the
		time evolution of gas above a certain density threshold, from $n_{\rm
			H}=10$ to $10^{6}\:{\rm cm^{-3}}$. The non-colliding case is
		represented by dashed lines, while the colliding case is represented
		by solid lines. For each model, $\sigma_{v}$ (averaged over the $x$,
		$y$, and $z$ lines-of-sight) for the total gas content (black),
		$\rm{CE_{2}}$ (blue), $\rm{CE_{3}}$ (green), $\rm{CE_{4}}$ (red), and
		$\rm{CE_{5}}$ (cyan) is plotted. }\label{fig:veldisp_times}
\end{figure*}

To understand the evolution of GMC kinematics in greater detail, we
analyze the velocity dispersion, $\sigma_{v}$, both as functions of
density (Figure~\ref{fig:veldisp_bins}) and time
(Figure~\ref{fig:veldisp_times}). $\sigma_{v}$ is calculated as the
cumulative mass-weighted gas velocity dispersion, taking the average
of the $x$, $y$, and $z$ lines of sight.  For the non-colliding model,
the components $\sigma_{v,x}$, $\sigma_{v,y}$, and $\sigma_{v,z}$
(not shown), are fairly similar throughout their evolution.  In the
colliding model, $\sigma_{v,x}$ is initially larger, but the
dispersions in fact converge by $\sim 2-3\:{\rm Myr}$ after the GMCs
have sufficiently interacted.

In Figure~\ref{fig:veldisp_bins}, $\sigma_{v}$ versus $n_{\rm H}$ is
plotted at various snapshots in time.  $\sigma_{v}$ for the total gas
and in CEs above given $n_{\rm H}$ thresholds are shown. The top row,
showing $t=1$, 2, and 3 Myr outputs, follows a similar trend in both
the non-colliding and colliding models.  As the GMCs evolve, the
$\sigma_{v}$ versus $n_{\rm H}$ relation reveals the formation of
higher density gas, with the colliding case exhibiting relatively
higher $\sigma_{v}$ throughout the entire density range due to the
bulk flows.  In both models, $\sigma_{v}$ versus $n_{\rm H}$ remains
fairly constant (i.e., a few ${\rm km\:s^{-1}}$) before dropping off
at higher $n_{\rm H}$.  Additionally, overall lower $\sigma_{v}$
values are measured in increasingly denser CEs.  At these earlier
times, the $\sigma_{v}$ at a given density is highest in the total
gas, revealing higher relative dispersions between sub-structures than
within sub-structures.

After $t\approx3\:{\rm Myr}$, this trend begins to shift, especially
in the colliding model. The GMC collision induces higher $\sigma_{v}$
for even the highest density gas. Higher-density ${\rm CE^{c}}$s
begin to have larger relative $\sigma_{v}$ as well.  At 4 Myr,
$\sigma_{v}$ within $\rm{CE_{2}^{c}}$, $\rm{CE_{3}^{c}}$, and
$\rm{CE_{4}^{c}}$ match that of the overall gas, while
$\rm{CE_{5}^{c}}$ has only slightly lower $\sigma_{v}$.  By 4.5 Myr,
$\sigma_{v}$ for the highest-density CEs actually exceed that of the
containing CEs and overall gas, indicating transition to a higher
velocity dispersion within dense structures versus between structures.
Conversely in the non-colliding model, the behavior at early times
remains, with lower $\sigma_{v}$ measured for higher-density CEs.

\citet{Offner_ea_2008} found in driven and decaying turbulent box 
simulations that turbulence driving produced structures with velocity 
dispersions similar or slightly higher in value within the densest regions 
(proto-stellar cores/clumps) compared to the core-to-core dispersion.
A sign of turbulent decay was that velocity dispersions within
cores sharply increased with density after a free-fall time.

Figure~\ref{fig:veldisp_times} follows the time evolution of
$\sigma_{v}$ for the total gas and in each CE.  Different panels
correspond to increasing gas density thresholds, above which
$\sigma_{v}$ is calculated.  Only CEs at the relevant density
threshold are plotted. However, $\sigma_{v}$ versus $t$ for a given CE
differs from panel to panel based on the density threshold.  For
example, consider $\rm{CE_{4}^{c}}$ in the $n_{\rm H} \geq
10^{5}\:{\rm cm^{-3}}$ panel (bottom center).  
The line associated with $\rm{CE_{4}^{c}}$ shows the velocity dispersion
for all gas $\geq 10^{5}\:{\rm cm^{-3}}$ within $\rm{CE_{4}^{c}}$, which
includes $\rm{CE_{5}^{c}}$ as well as other dense structures.
Such values can be used to compare overall dispersion not limited to a 
single CE.

\begin{figure}[htb]
	\begin{center}
		\includegraphics[width=1\columnwidth]{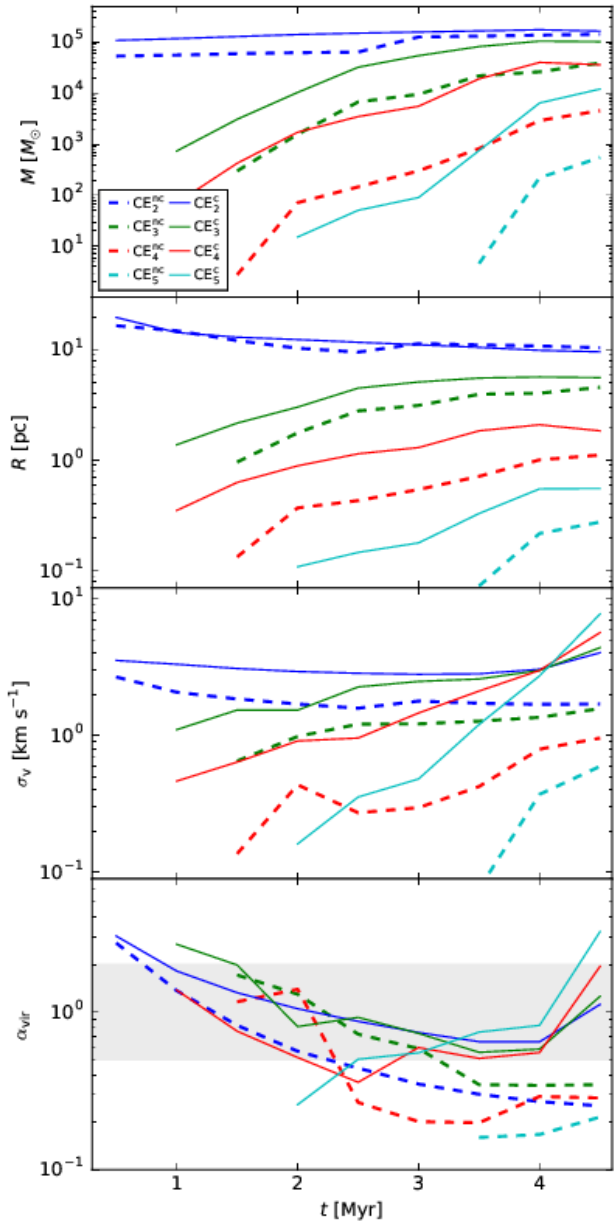} %
	\end{center}
	\caption{
		Virial analysis. (From top to bottom:) time evolution of total mass,
		half-mass radii, mean 1D velocity dispersion, and corresponding virial
		parameter. The non-colliding and colliding models are plotted as
		dashed 
		and solid lines, respectively. Values for respective CEs are shown in
		blue (${\rm CE_{2}}$), green (${\rm CE_{3}}$), red (${\rm CE_{4}}$),
		and cyan (${\rm CE_{5}}$).  }\label{fig:CE_veldisp_virial}
\end{figure}

The evolution of $\sigma_{v}$ for gas at lower-density thresholds (top
row) stays fairly constant at $\sim 3\:{\rm km\:s^{-1}}$ for the total
gas, $\rm{CE_{2}}$, and $\rm{CE_{3}}$ in both non-colliding and
colliding models.  For all thresholds and all extractions, the
colliding model produces higher $\sigma_{v}$ than the non-colliding
model by at least a factor of $\sim 2$.

For the $n_{\rm H} \geq 10^{4}\:{\rm cm^{-3}}$ threshold and higher,
the time evolution shows generally increasing $\sigma_{v}$.  For
higher density thresholds, $\sigma_{v}$ rises at a faster rate.
$t=4\:{\rm Myr}$ can be seen as the critical time in the colliding
case where $\sigma_{v}$ within structures exceeds that found
overall. At this time, all CEs are measured to have $\sigma_{v}
\approx 3\:{\rm km\:s^{-1}}$.  Structures formed in the non-colliding
model have a wide range of $\sigma_{v}$, from $\sim 1.5\:{\rm
  km\:s^{-1}}$ for $\rm{CE^{nc}_{2}}$ down to $\sim 0.4\:{\rm
  km\:s^{-1}}$ for $\rm{CE^{nc}_{5}}$.

\subsection{Dynamics}
\label{sec:dynamics}

Dynamical properties of gas structures can be described via a virial
analysis, which calculates the relative importance of a cloud's
internal kinetic energy to self-gravity.  This ratio of these energies
is defined as the virial parameter,
\begin{equation}
	\alpha_{\rm vir} = \frac{5 \sigma R}{G M},
\end{equation}
where $\sigma$ is the line-of-sight velocity dispersion, $R$ is the
radius, and $M$ is the total mass of the structure.  A uniform sphere
in virial equilibrium has $\alpha_{\rm vir} = 1$. Note, values of
$\alpha_{\rm vir}<2$ still imply a gravitationally bound structure if
magnetic and surface pressure terms are ignored \citep{Bertoldi_McKee_1992}.  
Virial analysis of a cloud may reveal signatures of its
kinematic history and thus can be a useful diagnostic to understand
dense gas formation triggered by GMC collisions. For example, a
collision and merger of two clouds would be expected to raise the
virial parameter of the combined structure, depending on how it is
defined.

Figure \ref{fig:CE_veldisp_virial} displays the virial parameter and
component properties measured for each CE.  Values for mass, radius,
velocity dispersion, and virial parameter are calculated for CEs in
the non-colliding and colliding GMC models.

In both models, the total masses of the CEs generally increase 
monotonically with time, as the extractions accrete surrounding gas 
and join with other structures. 
${\rm CE_{2}}$ masses formed in the collision are only slightly 
greater than those in the non-colliding case, as they track the initial 
GMC material. For higher density CEs, the relative difference grows up 
to factors of a few tens. Overall, the colliding case forms higher density 
CEs at approximately 0.5 to 1 Myr earlier times, with greater total masses.

The radius of a given CE is defined as the spherical radius calculated
from the mean CE gas density. The radii of ${\rm CE_{2}}$s decrease 
gradually as the GMCs contract gravitationally, while the higher-density CEs 
increase in size as additional material is included. While the non-colliding 
model retains a slightly larger measured radius relative to the compressed 
GMCs, the collision forms larger higher-density structures.

Larger differences are seen in velocity disperison, described in
detail in Section~\ref{sec:veldisp}.  Structures in the non-colliding
model have $\sigma_{v}$ that increase slowly with time throughout the
early evolution, with denser CEs exhibiting lower $\sigma_{v}$ for
the entirety of the simulation For colliding GMCs, structures with
much larger velocity dispersions are found. During early stages of
evolution, values of $\sigma_{v}$ are a factor of a few times higher
due to the contributions of the converging flows.  The velocity
dispersion increases at a more rapid pace for higher-density CEs,
with $\sigma_{v}$ for all CEs converging to roughly $3\:{\rm
  km\:s^{-1}}$ at $4\:{\rm Myr}$.  After this time, the highest
density CEs have higher velocity dispersions.
 
The combined influence of these parameters leads to notable
differences in the resulting virial parameter.  In both models, the
GMCs are initialized with moderately supervirial turbulence, which 
then naturally decays, as indicated by the initial ${\rm CE_{2}}$s.  
In the non-colliding case, the virial parameters of the ${\rm CE^{nc}}$s
decrease to sub-virial states by $t\sim 3\:{\rm Myr}$ and remain so for
the remainder of the simulation. Higher-density ${\rm CE^{nc}}$s are 
measured with lower $\alpha_{\rm vir}$, and all ${\rm CE^{nc}}$s reach 
$\alpha_{\rm vir}\simeq0.2-0.4$ by $t=3.5\:{\rm Myr}$ and remain fairly 
constant. Note the presence of large-scale $B$-fields can help stabilize 
these structures. The structures formed in the colliding GMCs have measured
virial parameters higher in general by a factor of a few, and remain near the
$\alpha_{\rm vir}\simeq0.5-2$ range from their formation to the end of the
simulation. Higher-density ${\rm CE^{c}}$s have higher 
$\alpha_{\rm vir}$--opposite of the trend seen in the non-colliding case. 
The virial parameters of all ${\rm CE^{c}}$s show growth after 
4 Myr, dominated by the increase in $\sigma_{v}$.

Observationally, \citet{Hernandez_Tan_2015} measured $\alpha_{\rm vir}
\sim 1$ for 10 IRDCs/GMCs (with dispersion of about a factor of 2) and
found that gas in IRDCs had moderately enhanced velocity dispersions
and virial parameters relative to GMCs, which may be indicative of
more disturbed gas kinematics particularly in denser regions.  This is
consistent with dense gas structures formed from our numerical
simulations of GMC collisions.

\subsection{Turbulent Momentum}
\label{sec:pdot}

\begin{figure*}[htb]
	\begin{center}
		\includegraphics[width=0.7\columnwidth]{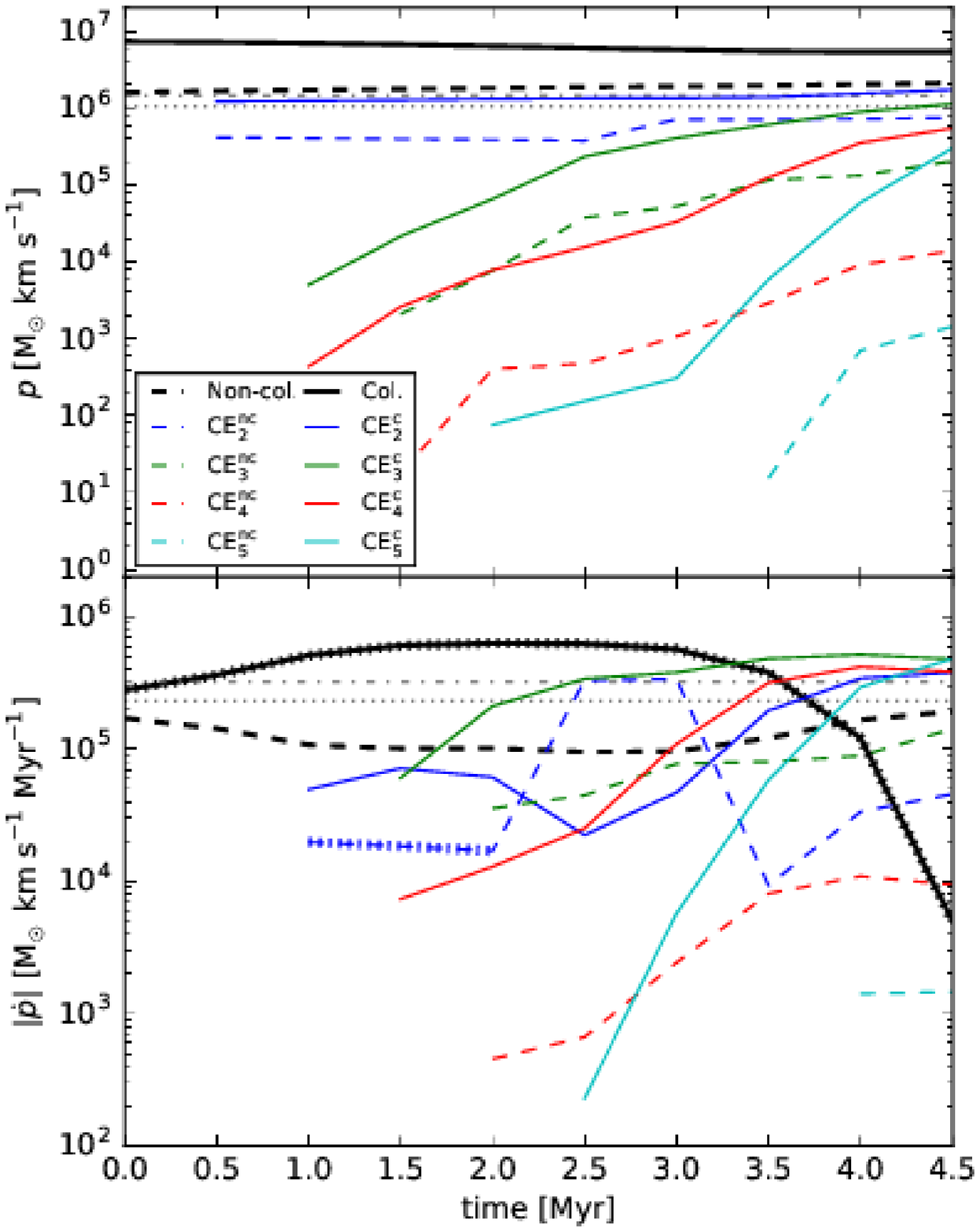}
		\includegraphics[width=1.29\columnwidth]{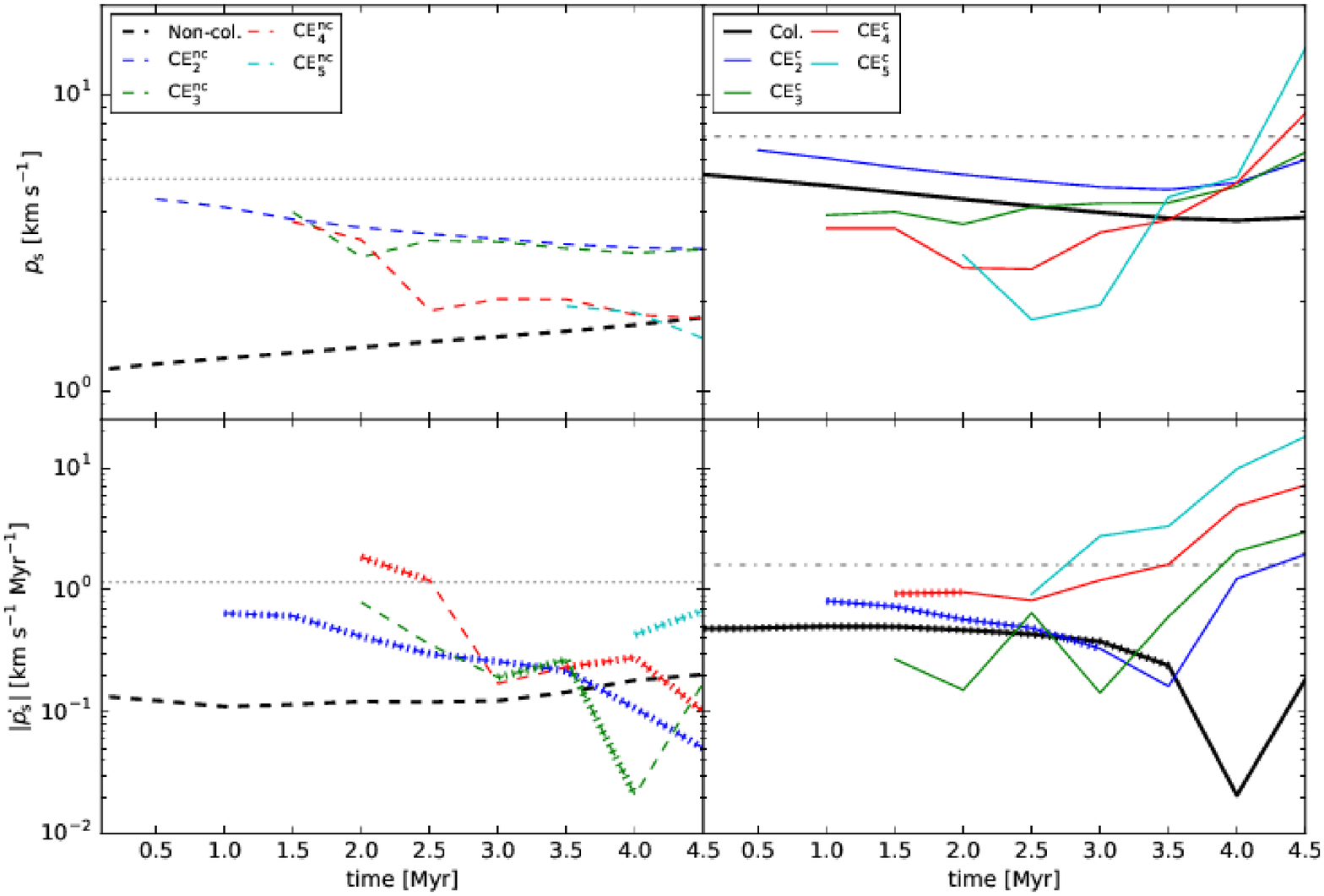}
	\end{center}
	\caption{
		(Left:) Time evolution of turbulent momentum (top) and 
		the magnitude of its time derivative (bottom), 
		with hatched segments denoting negative values.
		Total momentum is calculated for all gas 
		and individual CEs in the simulation frame. The non-colliding case is 
		drawn with dashed lines while the colliding case is drawn with solid lines. 
		For each plot, the grey horizontal lines indicate the expected initial 
		values for $p$ and the expected average $\dot{p}$ over 4.5 Myr for
		material contained within the
		non-colliding GMCs (dotted) and colliding GMCs (dash-dot), respectively. 
		(Right:) Time evolution of specific turbulent momentum (top), 
		$p_{\rm s}$, normalized to the total mass of the structure. 
		The absolute value of its time derivative (bottom), $|\dot{p}_{\rm s}|$ is
		also plotted, with hatched segments denoting negative values. 
		Non-colliding and colliding models are shown as dashed and solid 
		lines in separate columns, respectively, with the expected GMC initial 
		values in gray.
	}\label{fig:pdot}
\end{figure*}

The injection of turbulence from cloud collisions has been
investigated on galactic scales \citep{Tan_Shaske_Van_Loo_2013,Jin_ea_2017,Li_ea_2017arXiv}.  
From hydrodynamic simulations of flat rotation curve
galactic disks, \citet{Li_ea_2017arXiv} found that GMC collisions occurred
primarily with mass ratios only moderately less than 1, relative
velocities of $\sim$10s of ${\rm km\: s^{-1}}$, and timescales 
between subsequent collisions of
$\sim10$ to $20\:{\rm Myr}$ for conditions near a galactocentric
radius of $r\sim4~{\rm kpc}$.

The average rate of ``internal turbulent momentum'' injection per GMC,
$\dot{p}_{\rm CC}$, is expressed as
\begin{equation}
\label{eq:pdot}
\dot{p}_{\rm CC} = \frac{M_{\rm GMC} v_{\rm rel}}{t_{\rm col}}
\end{equation}
where $M_{\rm GMC}$ is the GMC mass, $v_{\rm rel}$ is the relative
collision velocity, and $t_{\rm col}$ is the average time between
collisions. Using the GMC collision parameters from \citet{Li_ea_2017arXiv} 
described above, $\dot{p} \sim 10 \times 10^{5}~M_{\odot}{\rm 
\:km\:s^{-1}\:Myr^{-1}}$.  Further, by balancing the total
injection and dissipation of turbulent momentum
leads to estimation of average GMC mass surface densities of 
$\Sigma_{\rm GMC, eq}\simeq 100\:M_{\odot}{\rm \: pc^{2}}$, 
consistent with observations.

Figure~\ref{fig:pdot} shows the time evolution for turbulent momentum,
	$p$ (top) and the absolute value of its time derivative, $|\dot{p}|$ (bottom).
	$p$ is calculated as the magnitude of the 
	total momentum for a given region, i.e., $p=\int_{V}{\rho |v| dV}$,
	a scalar quantity, with velocities calculated in the simulation frame. 
	$|\dot{p}|$ accounts for the total turbulent momentum injection as well as
	advection into the CE.
	The total momenta and rates are shown on the left, while quantities for the 
	specific momentum $p_{\rm s}$, i.e., normalized to the mass of the region 
	or CE, are displayed on the right. 
	Values for the total gas and all CEs are calculated for both the 
	non-colliding and colliding models.
	Also shown are horizontal lines displaying the expected initial value of $p$ 
	for the two GMCs in non-colliding and colliding models. The former accounts
	for the initial velocity dispersion of the GMC gas while the latter includes terms for
	the bulk colliding flows. Dividing the total 4.5 Myr gives the expected average rate,
	$\dot{p}$. By comparing CEs with this initial value, we can estimate the efficiency 
	of the conversion of turbulent momentum into denser structures.

For the total and ${\rm CE_{2}}$ gas, $p$ remains fairly constant over time,
	while higher-density CEs experience monotonic increases. $p$ in ${\rm CE^{nc}}$s 
	are in general smaller than in ${\rm CE^{c}}$s by up to a few orders of magnitude.
	Denser sub-regions have necessarily smaller $p$ but experience larger relative 
	growth which can be attributed to mass advection and, in primarily colliding cases, increased velocity dispersion. The ${\rm CE_{2}}$s, once both GMCs are taken into
	account, closely agree with the analytic expected initial values for $p$ within the
	GMC gas.

The rate at which this momentum is injected into the various structures, 
	$\dot{p}$, is also calculated. The non-colliding simulation domain is shown to 
	have slightly positive growth of turbulent momentum as material flows into the 
	volume. The colliding case shows negative values of $\dot{p}$, indicated by 
	the hatched lines, due to the lower velocity post-shocked ambient gas. 
	The individual CEs all exhibit positive $\dot{p}$ by 2~Myr, with ${\rm CE^{nc}}$s
	having generally lower rates ranging widely from $10^{3}-10^{5}~{\rm M_{\odot}\:km\:s^{-1}\:Myr^{-1}}$ and ${\rm CE^{c}}$s converging near 
	$4\times 10^{5}~{\rm M_{\odot}\:km\:s^{-1}\:Myr^{-1}}$. 
	Values for $\dot{p}$ for ${\rm CE^{c}_{2}}$ grow to those consistent with Equation~\ref{eq:pdot}, while ${\rm CE^{nc}_{2}}$ is generally an order of 
	magnitude lower.

To minimize the effects of mass advection into the structure, we investigate
	the specific momentum, $p_{\rm s}$, for each region. For the non-colliding 
	model, $p_{\rm s}$ gradually increases for overall gas but slowly decreases for 
	all ${\rm CE^{nc}}$s. This illustrates overall decay of turbulent momentum due to 
	decay of turbulence, though short-lived generation due to the collapse and attraction 
	of the clouds occurs. $p_{\rm s}$ of a few ${\rm km\:s^{-1}}$ are seen in this case.
	This general decrease is shown by ${\rm CE^{nc}}$s with 
	$\dot{p}_{\rm s}\approx-0.1{\rm km\:s^{-1}\:Myr^{-1}}$.

The $p_{\rm s}$ in the colliding GMC case decreases for overall gas as
	well as for ${\rm CE^{c}_{2}}$ initially. However, later stages and higher-density 
	${\rm CE^{c}}$s have $p_{\rm s}$ of $\sim 10~{\rm km\:s^{-1}}$. 
	Comparisons with the expected $p_{\rm s}$ show approximate turbulent momentum
	injection efficiencies of $\sim 0.7-0.8$ into ${\rm CE^{c}_{2}}$ for the majority of the 
	simulation and initially smaller efficiencies for higher-density ${\rm CE^{c}}$s, but 
	increasing over time. 
	The rates fluctuate as the collision occurs, but at later times results in
	${\rm CE^{c}_{2}}$ near $\dot{p}_{\rm s} \approx 1{\rm km\:s^{-1}\:Myr^{-1}}$ and 
	increased values for higher-density ${\rm CE^{c}}$s.  
	In contrast to the non-colliding case, turbulent momenta grows positively 
	in all the ${\rm CE^{c}}$s.

\section{Discussion and Conclusions}
\label{sec:conclusions}

We have analyzed three-dimensional turbulent MHD simulations of
non-colliding and colliding GMCs in order to investigate properties of
turbulence from GMC to clump scales and to determine whether or not
GMC collisions can be a source of supersonic turbulent momentum
injection.  Properties of the global gas and that which is contained
within increasingly dense, spatially-connected structures (CEs) are
compared. The morphology, volume density PDFs, velocity dispersion,
virial analysis, and rate of injection of turbulent momenta
are discussed.

The density PDFs reveal relatively smaller values of $\sigma$ in the
non-colliding case compared with the GMC collision.  The collision
produces gas which is more highly turbulent, with $\mathcal{M}\sim 13$
and increasing with time, while the more quiescent model has
$\mathcal{M}\sim 9$ and is decaying.

Velocity dispersions found in the total gas and especially within CE
structures were consistently higher in the colliding scenario.  Key
differences occur as the models evolve: in the colliding case, high
levels of velocity dispersion are maintained even at high densities
overall, whereas the non-colliding case forms high-density gas with
low velocity dispersions. At late stages, high-density ${\rm
  CE^{c}}$s have greater velocity dispersion than the low-density
${\rm CE^{c}}$s. In the non-colliding case, the opposite trend
occurs.

The virial parameter is also generally lower in the non-colliding
case, quickly decreasing and remaining at sub-virial states through
the course of the simulation. 
Note, however, that some support of the structures is 
	expected to be derived from the presence of large-scale 
	$B$-fields in the clouds.
The GMC collision, on the other hand,
forms approximately virialized dense gas structures that persist even 
into the late stages.

The rate of turbulent momentum injection, $\dot{p}$, was
found to be consistent with 
a simple model based on the initial global momenta of the 
	colliding clouds. This implies relatively high conversion efficiencies of 
	the bulk momenta of the collision into internal turbulence. These 
	results lend support to a model of maintenance of GMC internal 
	turbulence via GMC-GMC collisions that has been investigated in 
	simulations of shearing galactic disks by \citet{Li_ea_2017arXiv}.
We also note that the colliding GMCs produced
greater rates of specific turbulent momentum injection in higher density
structures whereas the non-colliding GMCs experienced 
decreasing specific turbulent momentum. 

In conclusion, parameters used to measure turbulence within dense gas
indicate that GMC collisions can indeed create and maintain states of
higher turbulence within dense gas structures. These are consistent
with various observations of dense IRDCs. Non-colliding GMCs, on the
other hand, experience turbulent decay with much less energetic
high-density structures.

\begin{ack}

Computations described in this work were performed using the 
publicly-available \texttt{Enzo} code (http://enzo-project.org). 
This research also made use of the yt-project (http://yt-project.org), 
a toolkit for analyzing and visualizing quantitative data \citep{Turk_ea_2011}. 
The authors acknowledge University of Florida Research Computing 
(www.rc.ufl.edu) and the Center for Computational Astrophysics (http://www.cfca.nao.ac.jp/)
for providing computational resources and support that have
contributed to the research results reported in this publication.
The authors thank the anonymous referee, whose comments helped to improve the article. 
BW acknowledges the JSPS Postdoctoral Fellowship (short-term) for its support.

\end{ack}

\end{document}